\documentclass[]{spie}
\usepackage[]{graphicx}

\title{On phase noise of self-injection locked semiconductor lasers}

\author{
E. Dale, W. Liang, D. Eliyahu, A. A. Savchenkov, V. S. Ilchenko, A. B. Matsko, D. Seidel, and L. Maleki
\skiplinehalf OEwaves Inc., 465 N. Halstead Street, Suite 140, Pasadena, California, 91107, USA}

\authorinfo{Send correspondence to A.B.Matsko:
Andrey.Matsko@oewaves.com}

\begin{document}

\maketitle

\begin{abstract}
We discuss our recent progress in iimproving the phase noise of a semiconductor laser using self-injection locking of  to a mode of a high-Q whispering gallery mode resonator. Locking efficiency is analyzed for semiconductor distributed feedback (DFB) as well as Fabry-Perot (FP) lasers operating at 690~nm, 1060~nm, 1550~nm, and 2~$\mu$m. Instantaneous linewidth below 300~Hz is realized with telecom DFB lasers. Tunability of the lasers is demonstrated. Commercially available packaged "plug-and-play" devices are manufactured.
\end{abstract}

\keywords{Lasers, Tunable Lasers, High-Q Resonators, Whispering Gallery Modes, Optical Phase Locked Loop \\ \\
{\small Reprinted from:
E. Dale ; W. Liang ; D. Eliyahu ; A. A. Savchenkov ; V. S. Ilchenko, et al. \\
``On phase noise of self-injection locked semiconductor lasers,'' \\
Proc. SPIE {\bf 8960}, Laser Resonators, Microresonators, and Beam Control XVI, 89600X (March 4, 2014); \\ doi:10.1117$/$12.2044824; $http://dx.doi.org/10.1117/12.2044824$ }
}

\section{Introduction}

Self-injection locking of semiconductor lasers \cite{dahmani87ol,hollberg88apl,himmerich94ao} using high quality factor (Q) whispering gallery mode (WGM) resonators allows realization of compact high performance devices. WGM resonators provide high-Q in a broad wavelength range \cite{savchenkovo7oe} and self-injection locking is one of the most efficient ways to lock a laser to a WGM. The locking effect occurs due to resonant Rayleigh scattering in the resonator \cite{gorodetsky00josab}, which results in back-reflection of some amount of light circulating in the resonator into the laser when the frequency of the light coincides with the frequency of the selected WGM. This provides a fast optical feedback resulting in laser linewidth collapse. Several experiments involving lasers locked to dielectric resonators, including WGM resonators, were reported previously \cite{vassiliev98oc,vassiliev03apb,kieu07ol,merrer08ptl,spengler09ol,sprenger10ol,llopis10ol,liang10ol,zhao11ol,peng11cpl,zhao12ptl,zhao12ol}.

Self-injection locking to a WGM is applicable to any laser emitting at a wavelength within the transparency window of the resonator host material. For instance, lasers emitting in the 1500~nm-10~$\mu$m range can be stabilized using CaF$_2$ or MgF$2$ WGM resonators. This leads to a broad range of opportunities for realizing miniature narrow-line lasers suitable for any application where low optical phase and frequency noise are important. As the self-injection locking does not require any electronics, the laser can be very tightly packaged, which simplifies its thermal management, and also reduces the influence of the acoustic noise on the laser frequency. The laser can be used as a master laser for pumping high power lasers used for metrology and remote sensing.

We realized self injection locking of DFB diode lasers using crystalline (CaF$_2$) WGM resonators \cite{liang10ol} and demonstrated instantaneous linewidth of less than 160~Hz in these lasers, with long frequency stability limited only by the thermal drift of the WGM frequency. The linewidth reduction factor determined as the ratio of the linewidth values of the free running and locked lasers was greater than 10,000. The minimal value of the Allan deviation for the laser frequency stability was $3\times 10^{-12}$ at the integration time of 20~$\mu$s.

Self-injection locking also facilitates production of tunable narrow-line semiconductor lasers \cite{hollberg88apl}. A narrow linewidth, widely tunable laser can be created from a thermally tunable diode laser injection locked to a tunable WGM. The tuning speed, however, is comparably slow in this case and is determined by the thermal response of the resonator fixture. The tuning rate is usually in the range of several gigahertz per degree. Agile frequency tuning of a self-injection locked laser can be realized by other means. For instance, electro-optic resonators with voltage controlled spectra can be utilized. The tuning agility in such a system is determined by the characteristic locking time, which can be shorter than a microsecond.

We reported earlier on packaging DFB diode lasers self-injection locked by means of electro-optically active WGM resonators into small form factor with 44$\times$27$\times$14~mm dimensions. We used lithium tantalate as the resonator host material and applied voltage to the resonator to tune the laser frequency \cite{ilchenko11spie}. In this work we report on demonstration of efficient optical phase locking of two electro-optically tunable self-injection locked lasers.

We also report on implementation of lasers with piezo-actuation and discuss improvements of noise parameters of packaged lasers operating in the telecom band. The piezo-actuation is advantageous since it allows locking the laser to a resonator made of any material, not necessary electro-optic. It is useful since non-electro-optic materials, such as MgF$_2$ and CaF$_2$, have much smaller optical loss, so the resonators made out of those materials usually have two to three orders of magnitude larger Q-factor. In turn, larger Q-factor results in tighter locking and narrower linewidth for the tunable lasers. In addition, usage of the non-electro-optic materials allows creating agile lasers in much broader wavelength range as compared to the lasers based on lithium niobate and lithium tantalate. We describe experimental results for lasers operating at wavelengths different from the telecom band and present data for the packaged devices operating at 690~nm, 1060~nm, 1550~nm, and 2~$\mu$m.

\section{Noise improvement achieved with self-injection locking}

\begin{figure}
\centering\includegraphics[width=14cm]{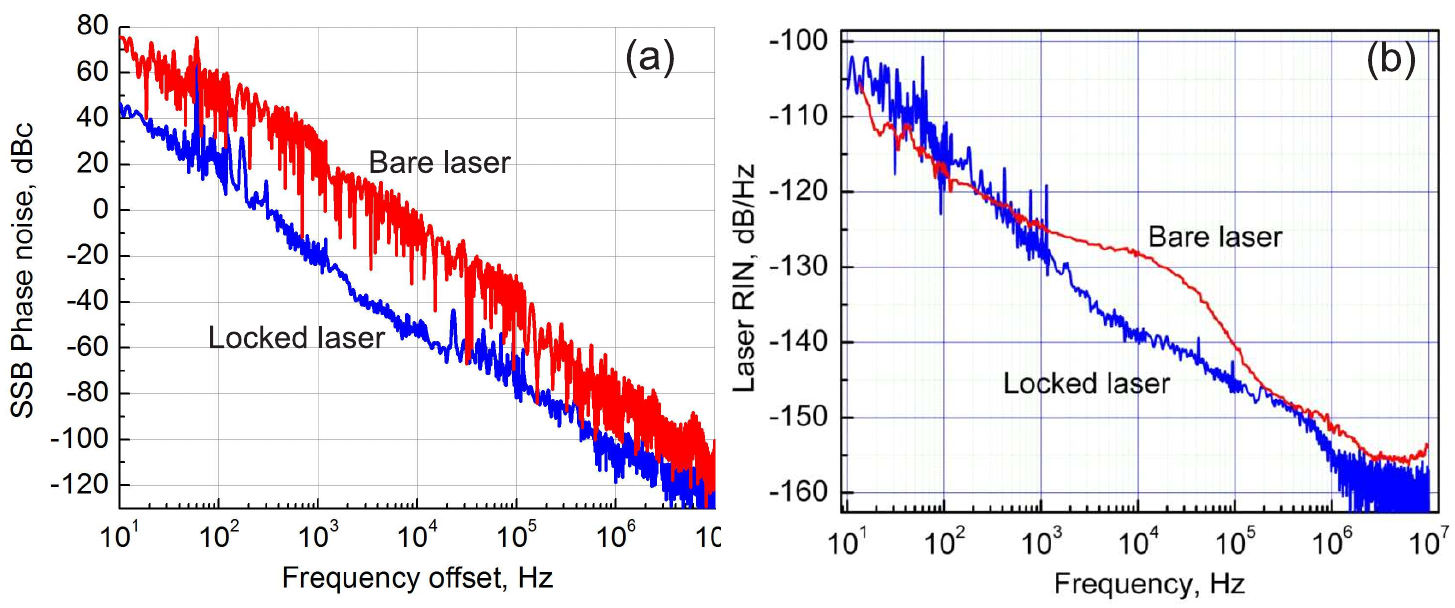}
\caption{\label{fig1noiseimprovement} Illustration of improvement of phase noise (a) and relative intensity noise (b) of a widely tunable semiconductor DFB laser due to self-injection locking to a mode of a high-Q WGM resonator.}
\end{figure}

We have studied self-injection locking of multiple types of semiconductor lasers and found that the process improve phase noise of the lasers rather significantly. A three orders of magnitude improvement of the laser phase noise, resulting from self-injection locking, can be seen at Fig.~(\ref{fig1noiseimprovement}). The phase noise of the locked lasers can be much smaller as compared with the noise shown in Fig.~(\ref{fig1noiseimprovement}) if resonators with higher quality factors are used to stabilize the lasers (see, e.g. phase noise of a self-injection locked laser in Fig.~\ref{fig1ultralow}).

The locking mechanism does not result in increase of the relative intensity noise of the laser (RIN). The noise even improves in certain spectral band (Fig.~\ref{fig1noiseimprovement}). RIN is sensitive to fluctuations of the feedback strength, which means that the optical path of the self-injection locked system has to be stable to avoid increase of the noise.

The power of self-injection locked lasers is usually limited because of nonlinearity of the high-Q resonators. The power of the light confined in the resonator mode should not cause stimulated Raman scattering and other types of instabilities. The lasers produced at OEwaves usually have output power of 10~dBm. As some applications require higher power levels, optical booster amplifiers can be used to increase the output power to a desirable level.

\begin{figure}
\centering\includegraphics[width=14cm]{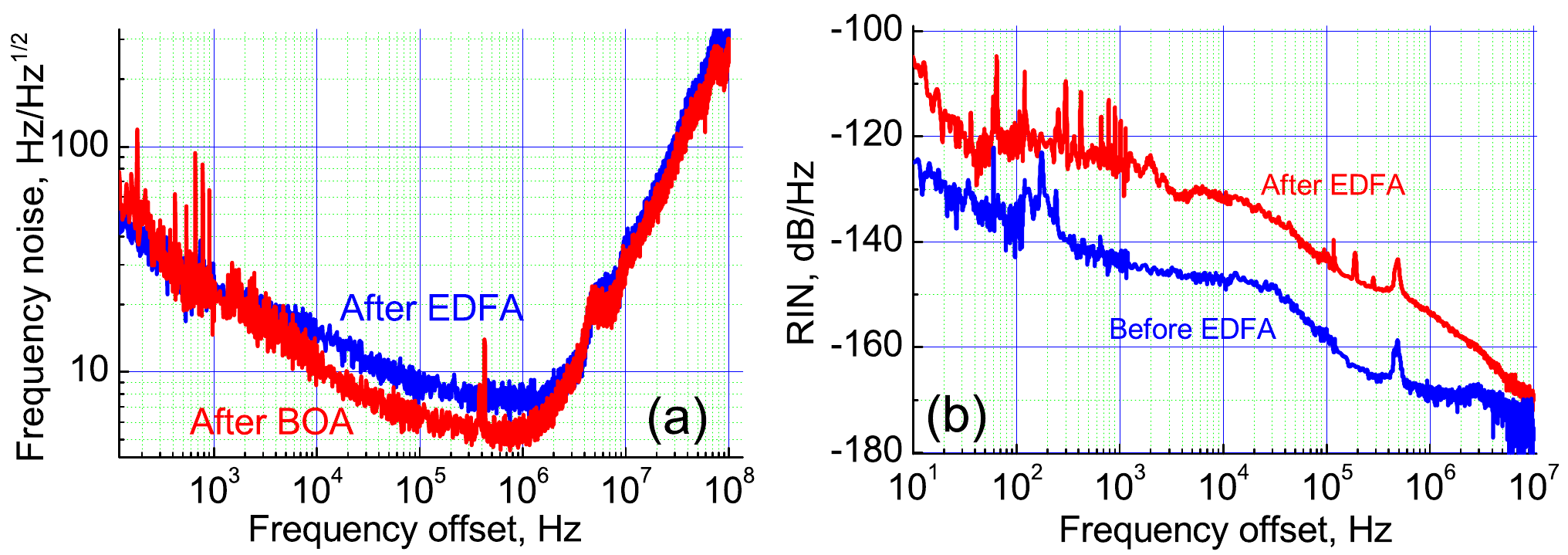}
\caption{\label{fig2noiseimprovement} Illustration of the modification of the laser frequency (a) as well as relative intensity (b) noise after amplification of the light emitted by a self-injection locked DFB laser. The output power is 16~dBm after the booster optical amplifier (BOA) and 20~dBm after the erbium doped fiber amplifier (EDFA). The laser output power is 6~dBm.}
\end{figure}

We verified the impact of optical amplifiers on the laser noise. The results of the measurements are presented in Fig.~(\ref{fig2noiseimprovement}). In this experiment we used a self-injection locked semiconductor laser with very low phase and frequency noise. The output power of the laser was approximately 6~dBm. We amplified the output light using either a semiconductor booster amplifier or an erbium doped amplifier and measured phase noise of the optical signal by beating the amplified light with optical local oscillator on a fast photodiode and measuring phase noise of the RF signal leaving the photodiode. The sensitivity of such a measurement is limited by the phase noise of the optical local oscillator. The measurement shows that the frequency noise of the amplified light is very good Fig.~(\ref{fig2noiseimprovement}a). The relative intensity noise degrades as compared with the noise of the base line laser output, however it is still rather good for the given value of optical power Fig.~(\ref{fig2noiseimprovement}b).

\section{Ultra-low noise 1550~nm lasers}

\begin{figure}
\centering\includegraphics[width=14cm]{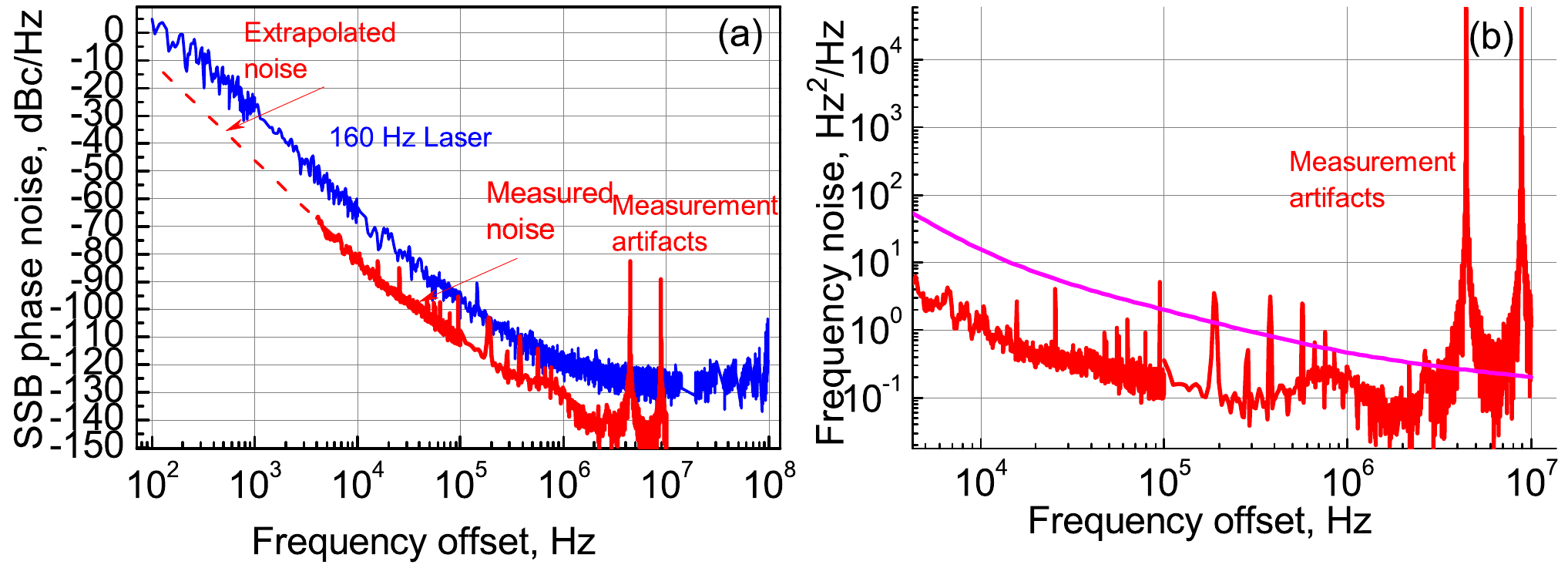}
\caption{\label{fig1ultralow} Demonstration of improvement of a C-band laser phase noise when a high-Q WGM is used for self-injection locking. (a) The figure show phase noise of the RF signal generated by beating two independent WGM-locked lasers on a fast PD (blue line, \cite{liang10ol}) versus phase noise of the laser assembled recently (red line). The 20~dB improvement in the phase noise is easily seen. The improvement in the phase noise corresponds to 10~dB improvement of the instantaneous linewidth. (b) The figure shows the frequency noise of the self-injection locked laser (red line) compared with the frequency noise of the low noise Brillouin laser (magenta line) reported very recently \cite{li14ol}.}
\end{figure}

We built a self-injection locked DFB laser prototype using high-Q MgF$_2$ resonator of 7~mm diameter. The intrinsic bandwidth of the resonator did not exceed 35~kHz, which means that the Q-factor was exceeding $5\times10^9$. The DFB laser was injected with 80~mA current and emitting 13~dBm of optical power. After the laser was self-injection locked to a mode of the resonator, we measured the phase noise using an OEwaves automated laser linewidth and frequency noise measurement system. The results of the measurements are shown in Fig.~(\ref{fig1ultralow}). The demonstrated laser has better phase noise as compared with the noise floor of the measurement system. An identical laser must be built to validate the predicted performance shown by the dashed line.

It is interesting to compare the performance of the self-injection locked laser with performance of a laser based on stimulated Brillouin scattering (SBS) \cite{li14ol}. SBS involves a nonlinear parametric interaction among the optical pump, Stokes, and acoustic modes, resulting in great suppression of the pump laser frequency noise. The fundamental Schawlow–Townes frequency noise of the SBS laser is on the order of 0.1~Hz$^2$/Hz. We achieve better performance with a conventional semiconductor laser locked to a high-Q microresonator. Self-injection locked laser can be integrated on a chip, similar to the SBS laser. The performance of the laser can be further improved if a higher-Q WGM resonator is used for injection locking.

\section{Wavelength variety}

\subsection{690~nm laser}

\begin{figure}
\centering\includegraphics[width=14cm]{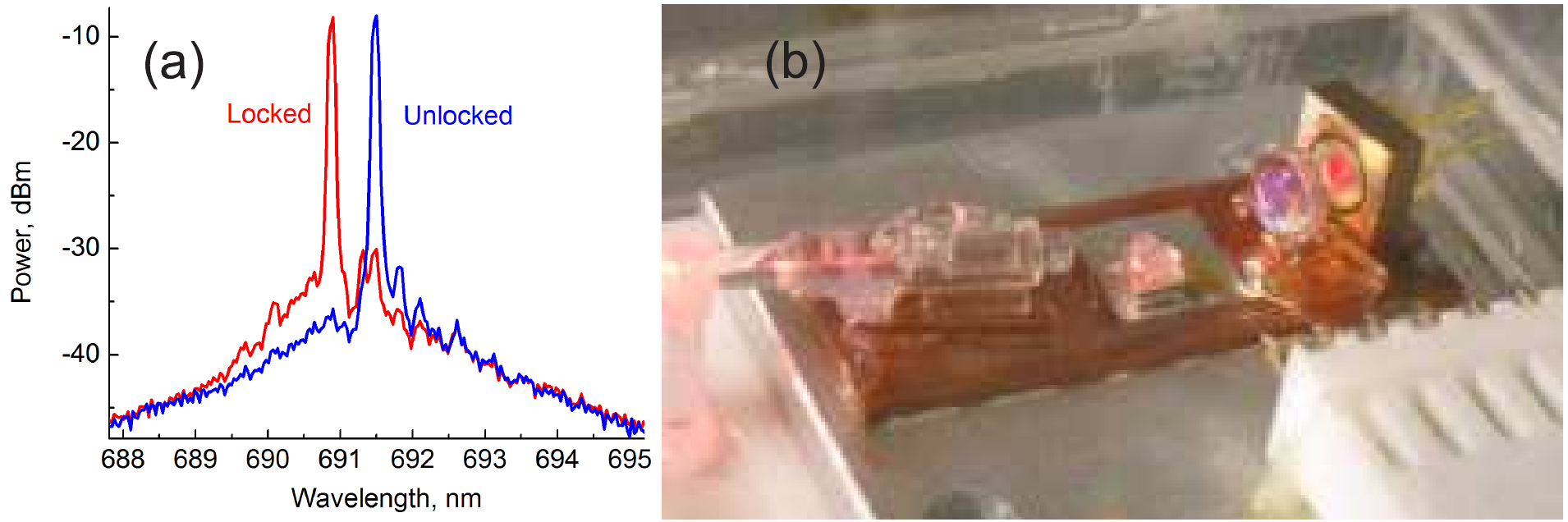}
\caption{\label{fig1FP690nm} Demonstration of self-injection locking of a Fabry-Perot semiconductor laser operating at 690~nm: (a) optical spectra of the free running and self-injection locked lasers; (b) a picture of the setup.}
\end{figure}

A commercially available single mode Fabry-Perot laser diode operating at 690~nm (Opnext 40~mW) was self injection locked to a high-Q WGM resonator (Fig.\ref{fig1FP690nm}).  The device was built on a monolithic optical bench with approximate dimensions of 0.5$\times$1 inches.  The frequency pulling effect of the self-injection locking process was far larger ($\approx$1nm) than self-injection locking with DFB ($\approx$0.1 nm) laser diodes due to the large difference in the optical quality factor of the WGM cavity and the Fabry-Perot cavity as well as mode competition with the laser diode.

Single mode Fabry-Perot laser diodes operate in a single frequency regime due to mode competition within the laser diode cavity.  The resonant optical feedback from the WGM resonator affects this mode competition and causes the laser diode to lase at an adjacent mode leading to large frequency changes when self-injection locking occurs.  This mode competition is also sensitive to changes in the feedback phase and magnitude which increases the sensitivity of the device to environmental perturbations when compared to self-injection locked DBR or DFB laser systems.

The self-injection locked FP laser system demonstrates a substantial decrease in RIN below 1~kHz when self-injection locked. This improvement should extend to much higher frequencies with addition of optical isolation before the output optical fiber.

\subsection{1060~nm laser}

\begin{figure}
\centering\includegraphics[width=14cm]{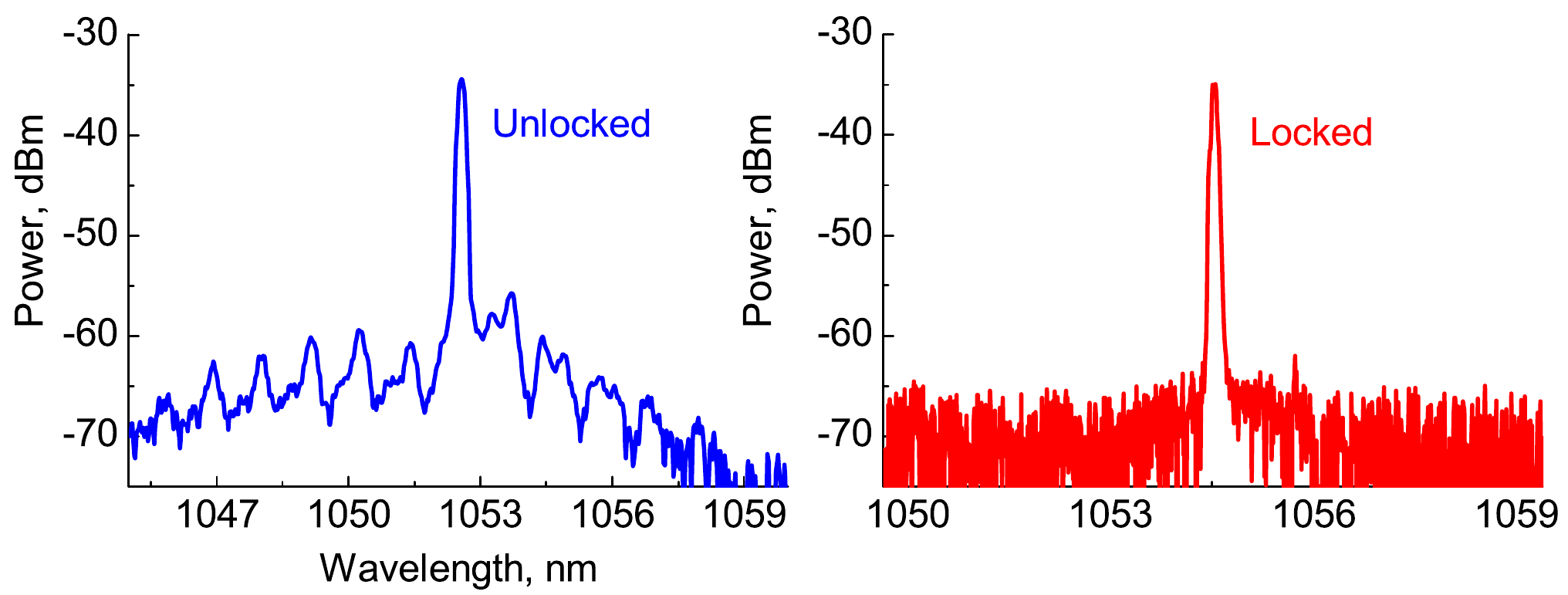}
\caption{\label{fig1FP1060nm} Demonstration of self-injection locking of a Fabry-Perot semiconductor laser operating at 1060~nm: (a) emission spectrum of a free running laser; (b) emission spectrum of a self-injection locked laser.}
\end{figure}

A bench top device was constructed from a single mode commercially available 1052~nm laser diode (Axcel Photonics 100~mW) and a high-Q whispering gallery mode resonator (Fig.~\ref{fig1FP1060nm}).  This device had similar injection locking characteristics to the 690~nm device showing large frequency pulling ($\approx$20~GHz) when locked.  The laser diode operated in single frequency operation when free running, but the mode competition was much stronger than that of 690~nm laser diode described above.  The strong mode completion causes reasonable suppression of the side modes. Self-injection locking results in further reduction of the side mode power.

\subsection{2~$\mu$m laser}

\begin{figure}
\centering\includegraphics[width=14cm]{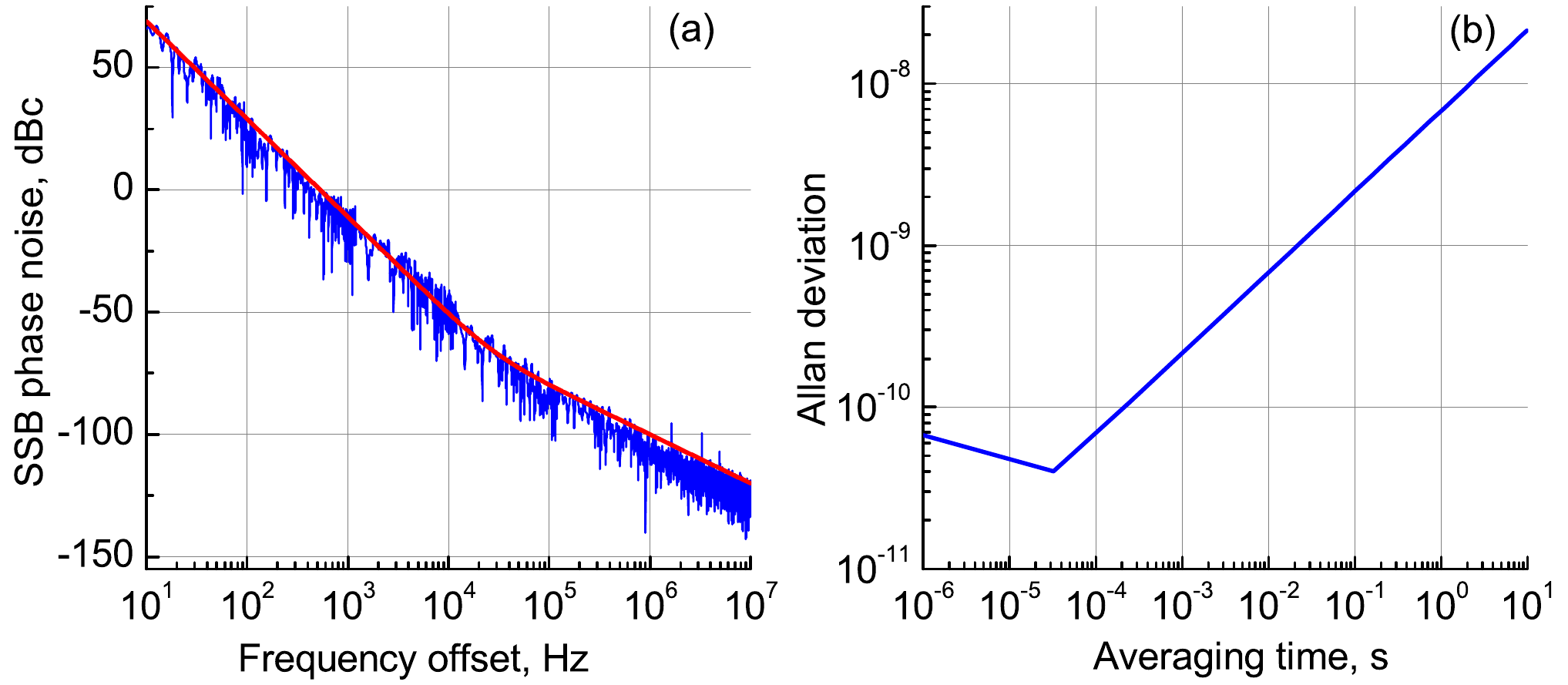}
\caption{\label{fig1DFB2um} (a) Results of a measurement of phase noise of a 2~$\mu$m laser (blue line) and a fitting curve (red line). (b) Allan deviation for the laser frequency calculated using the phase noise fitting curve.}
\end{figure}

A commercially available 2.05~$\mu$m DFB (Nanoplus, 18~mW) diode laser was self-injection locked to a high-Q WGM resonator with an integrated PZT (Fig.~\ref{fig1DFB2um}). The PZT allows high speed modulation of more than 10~MHz and flat frequency response for locking loops of more than 100~kHz.  This device was build into a custom fiber coupled package with dimensions of 3$\times$1$\times$1 inches which was then integrated into a turnkey consumer module for ease of use.  The dimensions of the completed laser module were approximately 6$\times$6$\times$1 inches.  The completed module had sub-kHz Lorentzian linewidth, more than 20~GHz of thermal tuning controlled via the input voltage on a single SMA port and approximately 2~mW output power from the fiber.  Status monitoring and laser condition are monitored via a USB connector on a PC with a custom control application.

\section{Electro-optical tuning}

\begin{figure}
\centering\includegraphics[width=14cm]{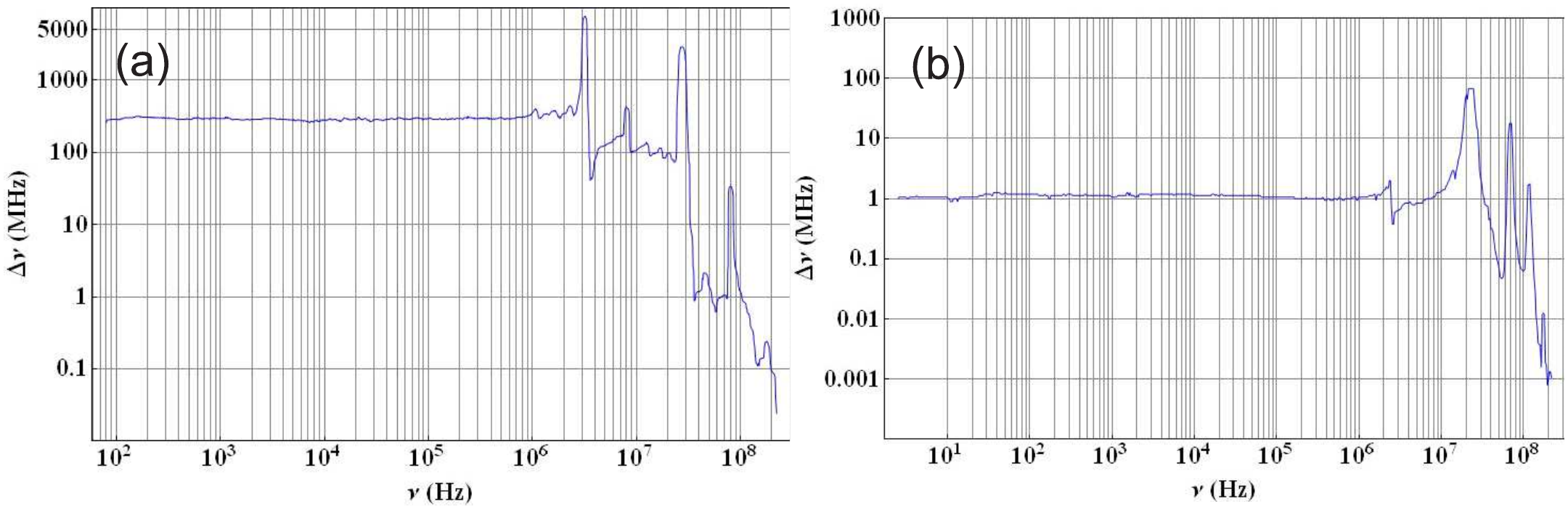}
\caption{\label{fig1EOtuning} (a) Demonstration of the electro-optical dithering for a laser self-injection locked to a mode of a lithium niobate whispering gallery mode resonator. The laser utput  power is 10~mW. Voltage applied to the resonator top and bottom faces changes by 10~V, peak to peak. This voltage change corresponds to 250~MHz frequency modulation span. The modulation frequency is limited by an internal electronic low pass filter with roll-off frequency of 5~MHz. (b) Demonstration of the laser dithering for the case of no electronic filter. The output power is 2~mW. The modulation span is 1.25~MHz.
 }
\end{figure}

We created an electro-optically tunable self-injection locked laser using a lithium niobate resonator (Fig.~\ref{fig1EOtuning}). Light emitted by a 1550~nm semiconductor DFB laser mounted on a ceramic submount was collimated and sent to a LiNbO$_3$ WGM resonator using a coupling prism. The power at the output of the laser chip was 25~mW. The maximal power at the output of the optical fiber was 10~mW.

The resonator was made of z-cut LiNbO$_3$ had 1~mm in diameter with unloaded Q-factor approaching $3\times 10^8$. Approximately 10\% of the light hitting the resonator mode was reflected back to the laser due to stimulated Rayleigh scattering,  locking the laser frequency to the WGM. In presence of optical feedback from the WGM resonator, the laser remained locked to WGM frequency within the range in excess of 2~GHz in terms of free running laser tuning.

The laser and the resonator were mounted on separate thermal control elements, and the long term stability of the laser were limited by residual temperature variations of the resonator. Another source of instability was related to intensity-dependent heating of the resonator perimeter, where WG modes are localized, due to absorption of light. Yet another source of frequency instability was the variations of the locking point of laser within the bandwidth of WGM due to residual variations feedback phase because of varying geometrical spacing between laser chip and resonator.

\subsection{Optical phase locking experiment with electro-optically tunable laser}

\begin{figure}
\centering\includegraphics[width=14cm]{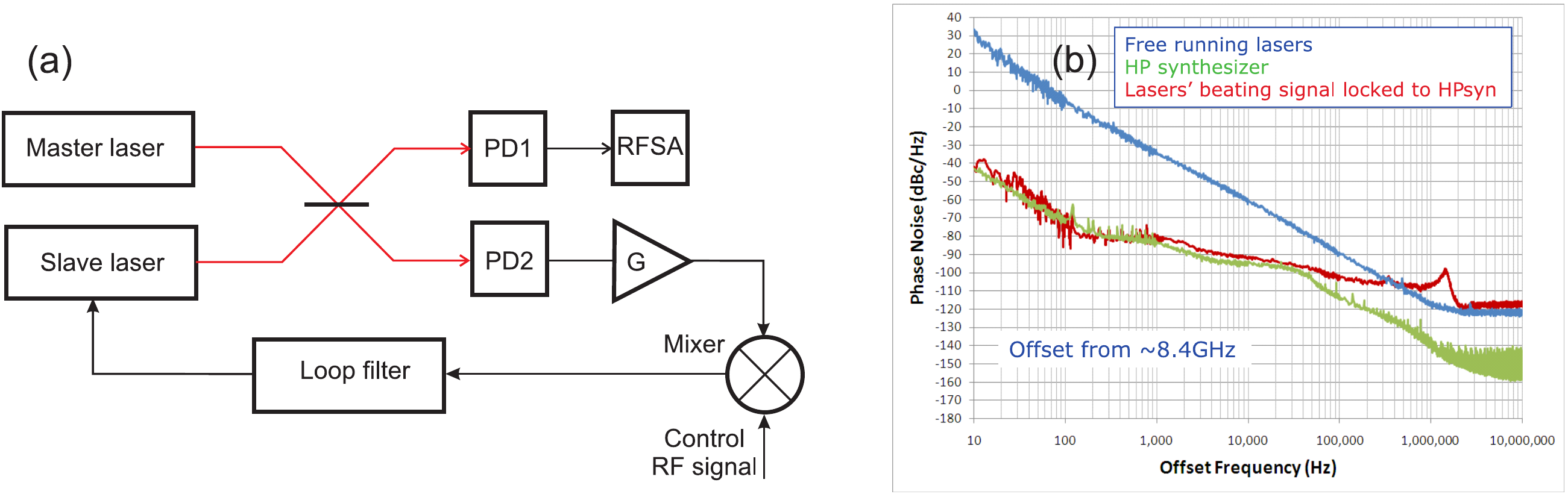}
\caption{\label{fig2EOtuning} (a) Schematic of the optical phase locked loop. (b) Phase noise modification as the result of locking the offset of two self-injection locked lasers to a HP RF synthesizer.
 }
\end{figure}

We built two narrow linewidth lasers by self-injection locking two DFB lasers to high-Q LiTaO$_3$ resonators respectively. Since LiTaO$_3$ is electro-optical material, whose refractive index can be changed quickly by electric field, we expected that these lasers could be modulated fast electrically. To demonstrate the fast modulation capability we then constructed an optical phase locked loop using the two narrow linewidth lasers as shown in Fig.~(\ref{fig2EOtuning}a). We mixed the two lasers’ signal on a fast photodiode and generated 8.4~GHz beat signal. By mixing the beat signal with a synthesizer’s signal, the phase different between the two lasers was extracted, filtered and fed back to the slave laser. In this way the frequency and the phase difference between the two lasers were locked to the synthesizer reference. The phase noise of the beat signal between the two lasers was also measured before and after engaging the phase lock loop, and compared with the phase noise of the synthesizer in Fig.~(\ref{fig2EOtuning}b). As can be seen, with the phase lock loop engaged, the phase noise of the beat signal, which represents the relative phase jitter between the two lasers, was following that of the synthesizer up to the bandwidth of the loop. The bandwidth of the loop was about 400~kHz and can be further increased to a few MHz if the loop filter is carefully designed and the loop delay can be shortened.

\section{Piezo-tunable laser}

\begin{figure}
\centering\includegraphics[width=14cm]{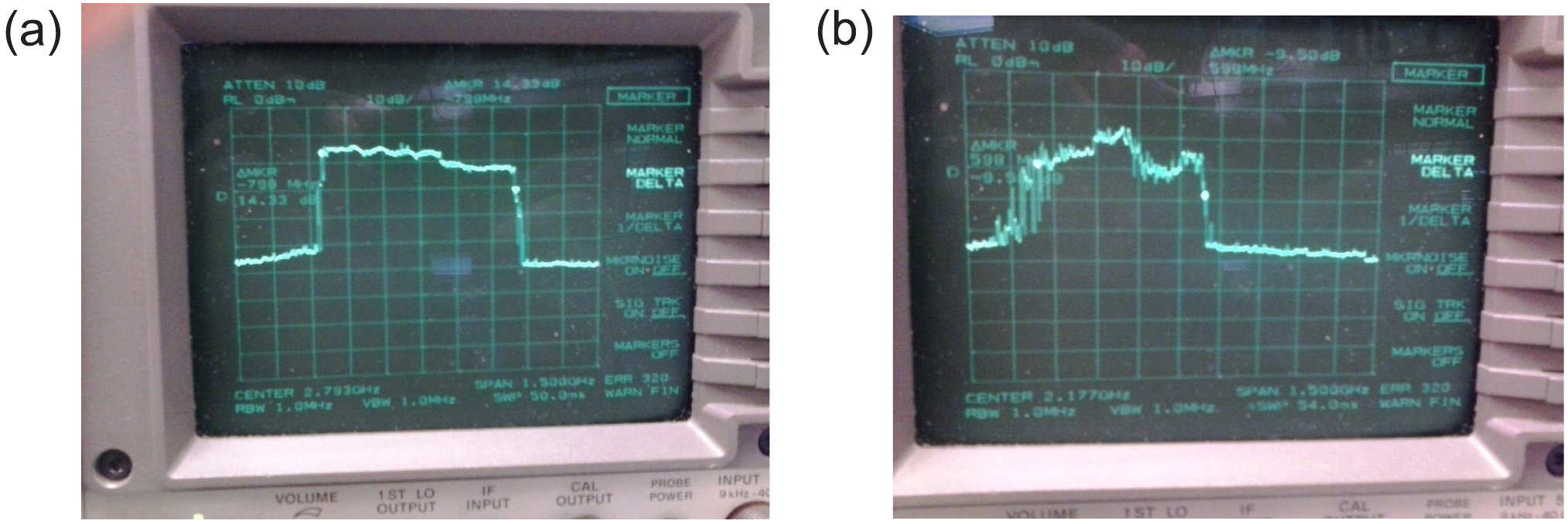}
\caption{\label{fig1piezotuning} Results of the tuning of frequency of a self-injection locked laser using a piezo-element attached to the WGM resonator. (a) Heterodyne beat note when the laser is dithered at 1~kHz frequency with 40~V peak to peak. The maximum frequency span is 800~MHz. (b) Heterodyne beat note taken when the laser is dithered at 100~kHz with 40~V peak to peak. The maximum frequency span is 600~MHz. Sensitivity and flatness of the response have reduced because the ILX current driver cut-off frequency is 100~kHz. A faster driver is required. }
\end{figure}

A standard OEwaves 1550~nm laser was equipped with a resonator with integrated PZT for high speed modulation and very fast loop bandwidth (Fig.~\ref{fig1piezotuning}).  The PZT stresses the microresonator through the elasto-optic effect creating a change in the local index of refraction at the optical mode.  The changing index causes the resonant frequency of the resonator to change. When self injection locked the laser tracks the changing frequency of the optical mode creating high speed frequency modulation in the optical output with very low residual amplitude noise.

The PZT-based self-injection locked laser was packaged in a fiber coupled butterfly package with a SMC microwave port connected to the integrated PZT. The package dimensions were approximately 1$\times$0.5$\times$0.5 inches.  The addition of the PZT resonator did not affect the performance characteristics of the standard laser, 50~GHz thermal mode hop free tuning, 2~nm complete wavelength coverage and 10~mW output power.  The size of the fully turnkey laser module was also unchanged with the addition of the PZT.  The high speed modulation and fast loop bandwidth can be used at any wavelength within the 2~nm of complete coverage.  Flat response range of more than 100~kHz has been demonstrated along with AC response frequencies demonstrated to be beyond 10~MHz.  The addition of the PZT to the WGM resonator does not change any of the optical characteristics of the resonator allowing this technique to be used to create high speed modulation at for injection locked lasers operating at any wavelength where the resonator has high-quality factor.
\begin{figure}
\centering\includegraphics[width=14cm]{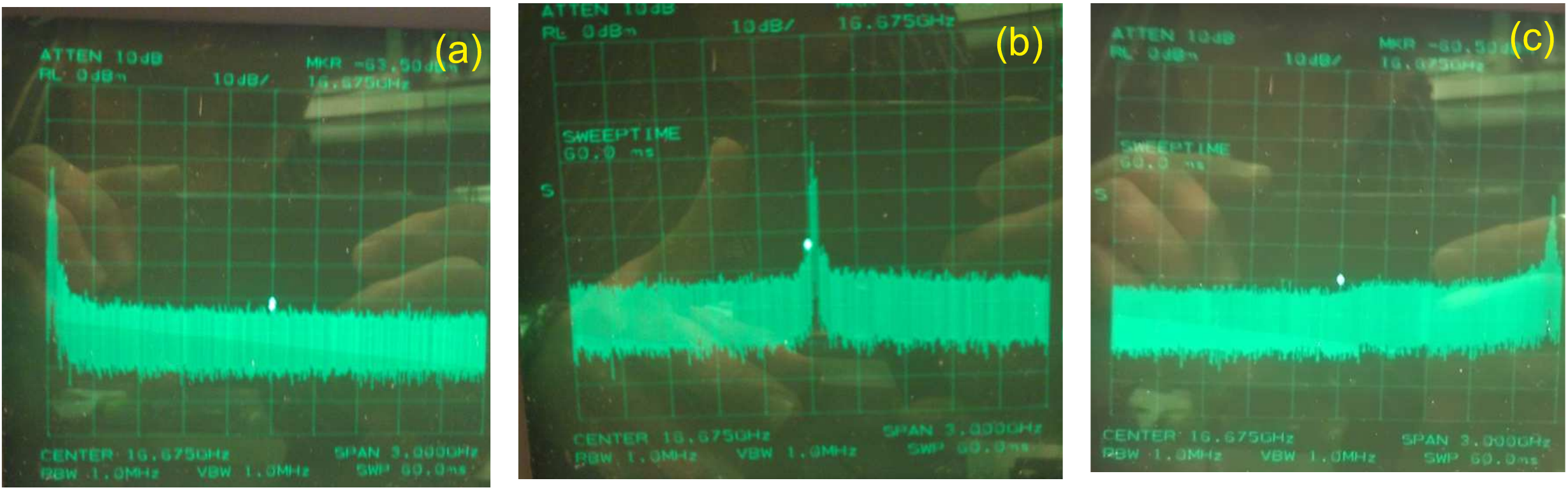}
\caption{\label{fig2piezotuning} Demonstration of DC tunability range for the laser. The tunability span is 3~GHz. (a) $V_{DC}=-117$~V; (b) $V_{DC}=0$~V; (c) $V_{DC}=90$~V.}
\end{figure}

\begin{figure}
\centering\includegraphics[width=14cm]{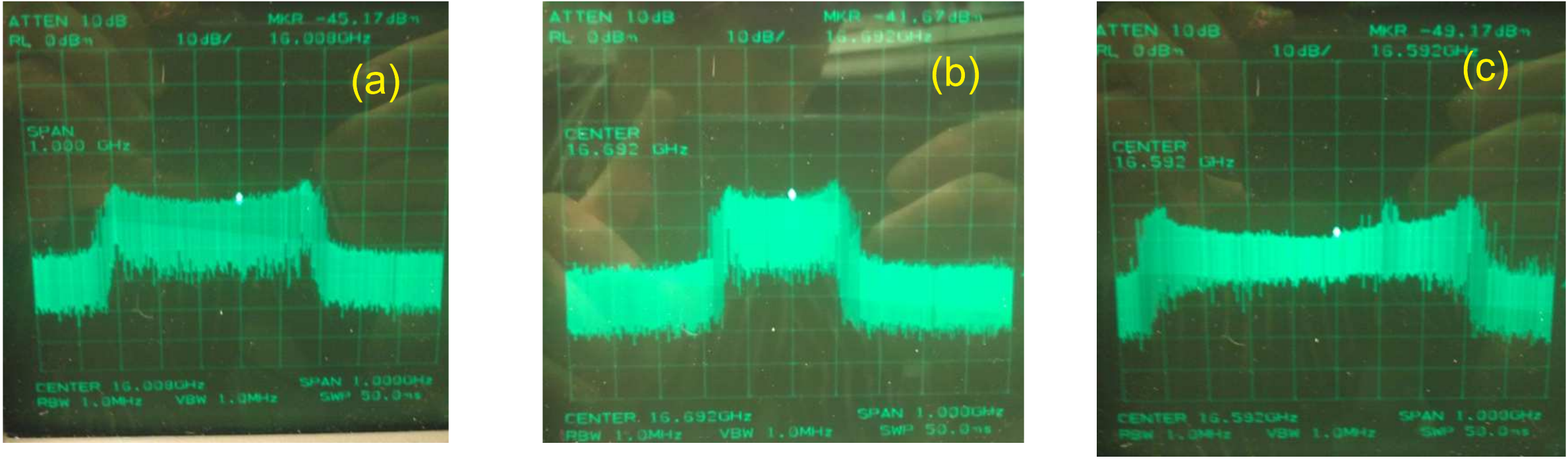}
\caption{\label{fig3piezotuning} AC frequency response of the self-injection locked PZT tunable laser. (a) Modulation frequency is 3~MHz; (a) Modulation frequency is 2~MHz; (a) Modulation frequency is 1~MHz. The laser is heterodyned using a fixed frequency laser and a small signal modulation is applied to the PZT.  Mechanical resonances are observed above 1~MHz.
}
\end{figure}

The performance of the laser tunable by means of a PZT actuator is illustrated by Figs.~(\ref{fig1piezotuning}), (\ref{fig2piezotuning}), and (\ref{fig3piezotuning}). The laser produces frequency modulated signal with very little admixture of amplitude modulation. The modulation frequency can exceed a MHz, and the modulation magnitude can exceed tens of MHz. Tunability span of 3~GHz is demonstrated.

\section{Conclusion}

We have studied self-injection locking of various lasers to modes of high-Q factor whispering gallery resonators. Fixed frequency as well as tunable lasers are demonstrated. The smallest measured absolute instantaneous linewidth of a 1550~nm DFB laser locked to a 50~kHz wide whispering gallery mode was 30~Hz. The laser has frequency noise of 0.1~Hz$^2$/Hz at frequency offset exceeding 10~kHz. The results reported here demonstrate the feasibility of the self-injection locking for fabrication of lasers in a very broad wavelength range limited only by the transparency range of the resonator host material.  Future work in this area will be focused on extending the range of WGM based injection locked lasers from uv to far IR.




\begin{thebibliography}{99}
\frenchspacing

\bibitem{dahmani87ol} B. Dahmani, L. Hollberg, and R. Drullinger, "Frequency stabilization of semiconductor lasers by resonant optical feedback," Opt. Lett. {\bf 12}, 876--878 (1987).

\bibitem{hollberg88apl} L. Hollberg and M. Ohtsu, "Modulatable narrow-linewidth semiconductor lasers," Appl. Phys. Lett. {\bf 53}, 944--946 (1988).

\bibitem{himmerich94ao} A. Hemmerich, C. Zimmermann, T. W. Ha\"nsch, "Compact source of coherent blue light," Appl. Opt. {\bf 33}, 988--991 (1994).

\bibitem{savchenkovo7oe} A. A. Savchenkov, A. B. Matsko, V. S. Ilchenko,  L. Maleki, "Optical resonators with ten million finesse," Opt. Express {\bf 15}, 6768--6773 (2007).

\bibitem{gorodetsky00josab} M. L. Gorodetsky, A. D. Pryamikov, and V. S. Ilchenko, "Rayleigh scattering in high-Q microspheres," J. Opt. Soc. Am. B {\bf 17}, 1051--1057 (2000).

\bibitem{vassiliev98oc} V. V. Vassiliev, V. L. Velichansky, V. S. Ilchenko, M. L. Gorodetsky, L. Hollberg, and A. V. Yarovitsky, "Narrow-line-width diode laser with a high-Q microsphere  resonator", Opt. Comm. {\bf 158}, 305--312 (1998).

\bibitem{vassiliev03apb} V. V. Vassiliev, S. M. Ilina, and V. L. Velichansky, "Diode laser coupled to a high-Q microcavity via a GRIN lens", Appl. Phys. B {\bf 76}, 521--523 (2003).

\bibitem{kieu07ol} K. Kieu, M. Mansuripur, "Fiber laser using a microsphere resonator as a feedback element," Opt. Lett. {\bf 32}, 244--246 (2007).

\bibitem{merrer08ptl} P. H. Merrer, O. Llopis, G. Cibiel, "Laser stabilization on a fiber ring resonator and application to RF filtering," IEEE Photon. Tech. Lett. {\bf 20}, 1399--1401 (2008).

\bibitem{spengler09ol} B. Sprenger, H. G. L. Schwefel,  L. J. Wang, "Whispering-gallery-mode-resonator-stabilized narrow-linewidth fiber loop laser," Opt. Lett. {\bf 34}, 3370--3372 (2009).

\bibitem{sprenger10ol} B. Sprenger, H. G. L. Schwefel, Z. H. Lu, S. Svitlov, L. J. Wang, "CaF$_2$ whispering-gallery-mode-resonator stabilized-narrow-linewidth laser," Opt. Lett. {\bf 35}, 2870--2872 (2010).

\bibitem{llopis10ol} O. Llopis, P. H. Merrer, A. Bouchier, K. Saleh, G. Cibiel "High-Q optical resonators: characterization and application to stabilization of lasers and high spectral purity microwave oscillators," SPIE {\bf 7579}, 7579--1B (2010).

\bibitem{liang10ol} W. Liang, V. S. Ilchenko, A. A. Savchenkov, A. B. Matsko, D. Seidel, and L. Maleki, "Whispering-gallery-mode-resonator-based ultranarrow linewidth external-cavity semiconductor laser," Opt. Lett. {\bf 35}, 2822--2824 (2010).

\bibitem{zhao11ol} Y. Zhao, Y. Peng, T. Yang, Y. Li, Q. Wang, F. Meng, J. Cao, Z. Fang, T. Li, and E. Zang, "External cavity diode laser with kilohertz linewidth by a monolithic folded Fabry–Perot cavity optical feedback," Opt. Lett. {\bf 36}, 34--36 (2011).

\bibitem{peng11cpl} Y. Peng, Y. Zhao, Y. Li, T. Yang, J.-P. Cao, Z.-J. Fang and E.-J. Zang, "Diode laser optically injected by resonance of a monolithic cavity," Chinese Phys. Lett. {\bf 28}, 114208 (2011).

\bibitem{zhao12ptl} Y. Zhao, Y. Li, Q. Wang, F. Meng, Y. Lin, S. Wang, B. Lin, S. Cao, J. Cao, Z. Fang, T. Li, E. Zang "A 100~Hz linewidth diode laser with external optical feedback." IEEE Photon. Technol. Lett. {\bf 24}, 1795--1798 (2012).

\bibitem{zhao12ol} Y. Zhao, Q. Wang, F. Meng, Y. Lin, S. Wang, Y. Li, B. Lin, S. Cao, J. Cao, Z. Fang, T. Li, and E. Zang, "High-finesse cavity external optical feedback DFB laser with hertz relative linewidth," Opt. Lett. {\bf 37}, 4729--4731 (2012).

\bibitem{ilchenko11spie} V. S. Ilchenko, E. Dale, W. Liang, J. Byrd, D. Eliyahu, A. A. Savchenkov, A. B. Matsko, D. Seidel, and L. Maleki, "Compact tunable kHz-linewidth semiconductor laser stabilized with a whispering-gallery mode microresonator," Proc. SPIE 79131G1--79131G9 (2011).

\bibitem{li14ol} J. Li, H. Lee, and K. J. Vahala, "Low-noise Brillouin laser on a chip at 1064~nm," Opt. Lett. {\bf 39}, 287-290 (2014).

\end{thebibliography}
\end{document}